\documentclass[a4paper,11pt]{article}
\usepackage{pos}

\title{Grid: OneCode and FourAPIs}
\ShortTitle{Grid: OneCode and FourAPIs}

\author[a,b]{Peter Boyle}
\author[c]{Guido Cossu}
\author[b]{Gianluca Filaci}
\author[d]{Christoph Lehner}
\author[b]{Antonin Portelli}
\author*[b]{Azusa Yamaguchi}

\affiliation[a]{Physics Department, Brookhaven National Laboratory, Upton, NY, USA}

\affiliation[b]{School of Physics and Astronomy,\\
University of Edinburgh, Edinburgh UK}

\affiliation[c]{Braid Technologies, Shibuya 2-24-12, Tokyo, Japan.}

\affiliation[d]{Universität Regensburg, Fakult\"at f\"ur Physik, D-93040, Regensburg, Germany}

\emailAdd{ayamaguc@staffmail.ed.ac.uk}

\abstract{We discuss a substantial update to the Grid software library for Lattice QCD,
  enabling it to port to multiple GPU architectures while retaining CPU vectorisation and
  SIMD execution within OpenMP threads. The GPU environments supported include vendor specific
  Nvidia CUDA and AMD HIP environments and a (mostly) standards based SYCL implementation.
  This is performed by an internal abstraction interface giving single source cross-platform
  performance portability across all number of planned Exascale architectures, and all those
  planned by the US Department of Energy.
}

\FullConference{%
 The 38th International Symposium on Lattice Field Theory, LATTICE2021
  26th-30th July, 2021
  Zoom/Gather@Massachusetts Institute of Technology
}

\begin{document}
\maketitle

\section{Introduction}

\begin{center}
\href{https://github.com/paboyle/Grid/}{https://github.com/paboyle/Grid/}
\end{center}

Grid is a software package that has for some time provided common operations and algorithms
that underpin many lattice QCD simulations \cite{GridManual,Boyle:2016lbp}.
It was initially developed\cite{Boyle:2016lbp} to support Single Instruction
Multiple Data (SIMD)
execution on a wide variety multicore processors by Intel, AMD, IBM and ARM, making use of SIMD intrinsics
to deliver excellent performance in compiled code.

Modern GPU's provide powerful alternatives to CPU's and deliver excellent performance and power performance for
a number of reasons. Firstly, an accelerator architecture may be more specialised than processor cores targetting best general purpose
single thread performance (and a full range of features). Secondly, by using a private memory system
more aggressive technology decisions may be made. A host CPU is retained for executing general code that is not well handled by
a GPU thread, and an \emph{offload} model is used where critical loops and data are marked by software for execution and placement
on the accelerator device.

All GPU's at this time present a parallel multi-dimensional (1d to 3d) loop as the primitive looping construct.
Similar to the Connection Machine computer, the machine model is that each instance of the parallel loop
body is presented as a different virtual machine or thread. However syntactically the implementation is less
elegant without data parallel expressions in the high level languge. Fortunately Grid provides such a high level
interface which is implemented on top of an internal parallel loop construct.

Although GPU's are fundamentally SIMD archictures, addressing modes and masked execution are cleverly used to obscure this
fact and present a scalar processing model to the programmer, called ``Single Instruction Multiple Thread'' (SIMT).
In SIMT, a single instruction fetch unit controls \emph{multiple} logical threads, typically a number of O(32).
When some threads choose yes, and other threads choose no the divergence leads to loss of parallel throughput.

Accesses to thread private data (stack, local memory and what the programmer would think of as local variables - if not in registers)
are addressed in a way that efficiently interleaves accesses to corresponding local memory locations by each thread in
a physical memory array.
One might imagine that electronically the ``thread'' index withing a parallel execution group dictates the byte address within
a hardware SRAM data bus. This ensures that when a group of software ``threads'' concurrently execute the same instruction, and they
all access the matching variable on their respective stack or local memory. The accesses will be transferred as a physically contiguous
data beat even though the virtual addresses are relative to a stack pointer or in a local memory space.
\footnote{One might even imagine that conventional microprocessors could, in principle, add addressing modes that facilitate a similar SIMT model in their
vector extensions.}

The challenge of writing high performance and portable code is three fold.
Firstly, the syntax for offloading loops depends on the underlying software environment.
We have managed to identify a suite of abstractions that are both compact and adequate to write portable
and performant software with a single interface.
Secondly, the somewhat larger challenge is to write a single programme that captures the differing semantics between SIMD
and SIMT execution models. Our goal is to preserve 
Thirdly placing data and managing data motion should be simple and even transparent when using Grid data parallel operations.

In this proceedings we highlight the changes made to Grid to obtain cross-platform accelerator portability and we
give some early performance results on modern variety of architectures. We demonstrate that the code is provably optimal
on Nvidia A100 GPUs for the Domain Wall Fermion operator.

\section{Acceleration abstraction}

The generalisation of Grid to GPU's required us to introduce several related technologies.
The target is to define an abstraction that can cover HIP, SYCL and CUDA

\begin{itemize}
\item Offload primitives and device function attributes
\item Memory allocation primitives
\item Software managed device cache for host memory regions
\item Distinguish accessors (views) of lattice objects from the storage container
\item Abstraction capturing SIMT and SIMD models in a single interface
\item Updating the Grid Expression Template engine
  \end{itemize}

Grid already had a parallel for construct used to target OpenMP threaded loops on multicore
CPUs. This was generalised, using a similar C++ Lambda function object 
approach to that taken by SYCL\cite{SYCL}, Kokkos\cite{Kokkos} and RAJA\cite{Raja} to
capture loop bodies and pass to a device. Care must be taken to ensure all data referenced
by the loop body is accessible to the device and we will describe how this is performed later.

{\bf Covariant programming:}
The optimal data layout changes with parallelism model. Both SIMD and SIMT are electronically vector architectures
  and a partial ``struct-of-array'' transformation is needed in data arrays in memory.
  However they semantically differ in the behaviour of local variables within functions.
  In GPU each ``lane'' of the underlying SIMD executes a different logical instance of the same function, and thus processes
  scalar items, while in a CPU local variables remain (short) vector data types.
  Optimal software cannot be invariant when the architecture is changed, and rather
  to target both efficiently it is necessary to design a programming style that transforms covariantly with the architecture, as in the
  table below.

\begin{tabular}{|c|c|c|}
\hline
Model   & Memory  &  Thread \\
\hline
Scalar  & Complex Spinor[4][3]         & Complex Spinor[4][3]  \\
\hline
SIMD    & Complex Spinor[4][3][N]      & Complex Spinor[4][3][N] \\
\hline
SIMT    & Complex Spinor[4][3][N]      & Complex Spinor[4][3] \\
\hline
\end{tabular}

Grid introduces transfer functions coalescedRead and coalescedWrite that map between these
layouts, and lattice objects have an additional accessor method \verb1 operator() 1 that performs
this translation and minimises the syntactical changes between SIMD and SIMT loop bodies.
C++11 added automatic type inference and the key to \emph{covariant programming} is to not hard code
the datatypes of temporary variables in a loop, but to deduce the type from the return type
of a coalescedRead so that the loop body transforms with the architecture. We give an illustrative example
below of an opmitised routine and explain the elements in the following sections.

{\small
\begin{verbatim}
template<class obj1,class obj2,class obj3> inline
void mult(Lattice<obj1> &ret,const Lattice<obj2> &lhs,const Lattice<obj3> &rhs){
  ret.Checkerboard() = lhs.Checkerboard();
  autoView( ret_v , ret, AcceleratorWrite);
  autoView( lhs_v , lhs, AcceleratorRead);
  autoView( rhs_v , rhs, AcceleratorRead);
  accelerator_for(ss,lhs_v.size(),obj1::Nsimd(),{
    decltype(coalescedRead(obj1())) tmp;
    auto lhs_t = lhs_v(ss);
    auto rhs_t = rhs_v(ss);
    mult(&tmp,&lhs_t,&rhs_t);
    coalescedWrite(ret_v[ss],tmp);
  });
}
\end{verbatim}
}

\subsection{Internal API}

The functionality of Grid was augmented with ``accelerator'' primitives.
\begin{itemize}
\item Function attributes
{\small  
\begin{verbatim}  
    accelerator
    accelerator_inline
\end{verbatim}
}
\item Parallel looping / offload
{\small  
\begin{verbatim}
    accelerator_for(iter1, num1, nsimd, ... )
    accelerator_for2d(iter1, num1, iter2, num2, nsimd, ... )
    accelerator_forNB, accelerator_for2dNB 
    uint32_t accelerator_barrier();         // device synchronise
\end{verbatim}
}
\item Parallelism control: Number of threads in thread block is acceleratorThreads*Nsimd
{\small  
\begin{verbatim}
    acceleratorInit();
    uint32_t acceleratorThreads(void);   
    void     acceleratorThreads(uint32_t);
    void     acceleratorSynchronise(void);   // synch warp etc..
\end{verbatim}
}
\item Coalesced reading support
{\small  
\begin{verbatim}
    int      acceleratorSIMTlane(int Nsimd); // my thread location
    // Memory representation to stack representation
    coalescedRead()/coalescedReadPermute()/coalescedWrite()   
\end{verbatim}
}
\item  Reduction
{\small  
\begin{verbatim}
    template<class t> accelerator_sum(t *tp,uint64_t num)
\end{verbatim}
}
\item  Memory management and motion
{\small  
\begin{verbatim}
    void *acceleratorAllocShared(size_t bytes);
    void *acceleratorAllocDevice(size_t bytes);
    void acceleratorFreeShared(void *ptr);
    void acceleratorFreeDevice(void *ptr);
    void *acceleratorCopyToDevice(void *from,void *to,size_t bytes);
    void *acceleratorCopyFromDevice(void *from,void *to,size_t bytes);
    void *acceleratorCopyDeviceToDevice(void *from,void *to,size_t bytes);
\end{verbatim}
}
\end{itemize}

\subsection{Offload primitives and attributes}

Grids internal API to acceleration is contained in a header \verb1 Accelerator.h 1, and is itself a fairly useful component.
Generically the prefix \verb1 accelerator 1 is used in the functionality.
CUDA requires by default that device code be in ``.cu'' source files. This can be avoided with compiler flags to
insist that all C++ files contain CUDA code, and not renaming.
CUDA and HIP have a compiler model that requires that all accelerator functions be marked with a \verb1 __device__ 1 attribute.

\begin{figure}[hbt]
  \includegraphics[width=\textwidth]{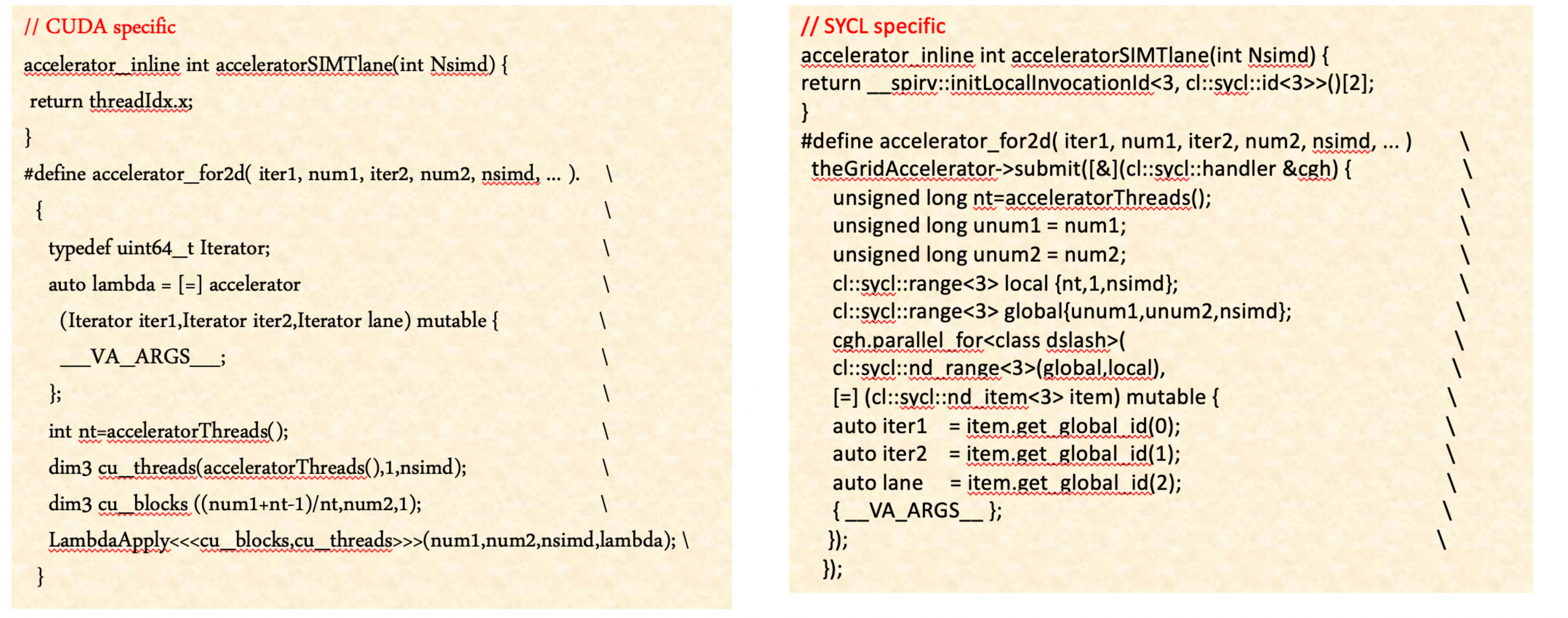}
  \caption{\label{fig_cuda_sycl}
    Macro implementation of kernel offload for CUDA and SYCL.
    Grid and user code use consistently the accelerator\_for construct.
    We emphasise that most user code uses either Grid functions or expression template engine and only expert kernels
    use the accelerator\_for . This is an internal implementation detail that may be useful to others developing independent GPU codes.
  }
\end{figure}

\begin{figure}[hbt]
  \includegraphics[width=\textwidth]{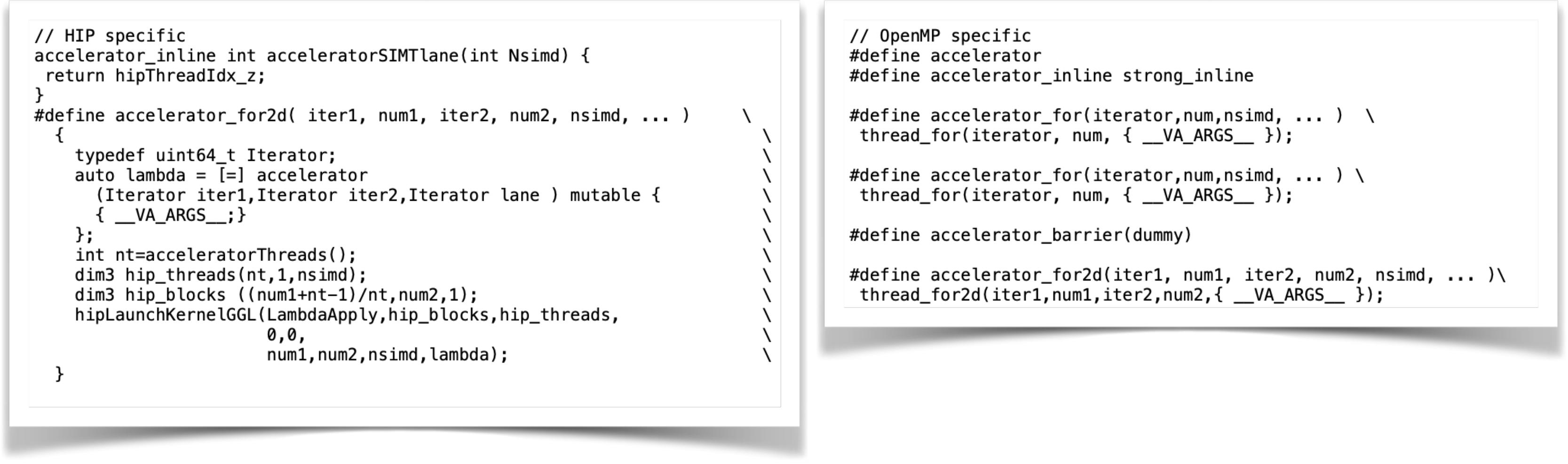}
  \caption{\label{fig_hip_openmp}
    Macro implementation of kernel offload for HIP and OpenMP.
    Grid and user code use consistently the accelerator\_for construct.
    We emphasise that most user code uses either Grid functions or expression template engine and only expert kernels
    use the accelerator\_for. This is an internal implementation detail that may be useful to others developing independent GPU codes.
  }
  \end{figure}

For Grid code itself, handling this is not onerous: Grid has always used a \verb1 strong_inline 1 attribute for high performance
code, and globally renaming this attribute \verb1 accelerator_inline 1 as a macro that on HIP and CUDA expands to give both device and inline attributes.
The \verb1 parallel_for 1 construct was replaced with distinguished \verb1 thread_for 1 which always executes on the host processor under OpenMP
and \verb1 accelerator_for 1. These macros capture a loop body as a macro parameter and on GPU targets form a hidden C++ lambda function object
that executes one loop iteration. The object is passed as a \verb1 device lambda 1 on HIP, SYCL and CUDA.
Examples of the macro implementation are shown in figure~\ref{fig_cuda_sycl} and \ref{fig_hip_openmp}

\subsection{Memory models and software managed cache}

Grid can be compiled with two options for using GPU memory. The simplest (and earliest port)
was achieved using Unified Virtual Memory (UVM) where we assume that memory can be allocated for lattice (and other)
data that is accessible to both CPU code and the accelerator loops.

This was found to perform reasonably well until, particularly on the Summit computer, the total capacity of the
GPU memory was exceeded and substantial slow down was seen when data had to be evicted to make space for new data.

As a result we also implemented a MemoryManager object that maintains a software cache of host memory
on the device with a replacement algorithm under our control. This is inspired to some degree by the SYCL ``buffer''
model but leaves code able to use pointers.

A key element is to separate Lattice objects into \emph{lattice containers} which own the data, and \emph{lattice views}
which contain pointers and the ability to dereference or access the data. A view is obtained by calling a member function of the
container. The view is a lightweight structure appropriate to be copied by value into a device kernel.
The call to obtain the view must specify intent: one of CpuRead, CpuWrite, AcceleratorRead or AcceleratorWrite.
Under UVM compilation (--enable-unified=yes) the operation is trivial.
However under explicit data motion (--enable-unified=no) this allows a software cache to be consistently maintained.

The MemoryManager contains two data structures: a table of cache entries, indexed by host pointer and storing
(possible) corresponding device pointer, region size, reference counters and
a state that is one of CpuDirty, AcceleratorDirty, Consistent or Empty.
The sequence of view accesses migrates a vector between host and device according to access intent and prior
state. The whole buffer is treated as a single entity and high performance memory copies between host and device used.
The total aggregate footprint available to Grid for this cache has a target high watermark limit, controllable
via a command line parameter --device-mem X (mb). If this high watermark will be exceeded by moving data to the
device, previously resident data is evicted to make space.

The cache data structure is implemented via an O(1) overhead hash table (std::unordered\_map)
The victim selection implements a true ``least recently used'' algorithm. The LRU is maintained using
an O(1) double ended queue to maintain ordering. Any access removes an item from the queue and replaces
it at the front.
As a power user feature,  Lattice fields can be given one of two priories with an ``Advise'' function.
Large volume and infrequent data can be advised as infrequently used, and made always a higher priority

Views are opened locally in a scope around an accelerator\_for loop, and opening and closing 
the (reference counted) Views trigger the software cache operations. A convenience ``autoView'' macro hides some
of the syntactical overhead for closing.
View management is automatic when using the Grid Expression template engine.

\subsection{Expression template engine}

Grid has for some time had a flexible expression template engine. Updating this to
work efficiently under offload required some careful implementation.
Grid builds a compound object representing the abstract syntax tree (AST) of an expression,
and a deferred evaluation function performs the operations that this AST represents.
The composite object built has to no longer store references to the lattice containers
(these would be host pointers!) but rather map these to accelerator read view objects.
The evalution of lattice leaf nodes in this expression tree were updated to return a scalar
element, the result of a coalesced read on the lattice object.
Although very sophisticated and carefully written C++ code, the modifications are actually rather modest
and general expression template user code works without modification.

\section{Performance results}

{\bf CUDA}
The Univeristy of Edinburgh and Juelich Supercomputer Centre have both recently purchased ATOS systems based
on nodes with four A100 Nvidia GPUs, AMD CPU's and four Mellanox HDR network interfaces. The system uses PCI express
switches and gives good bus performance between network and GPU memory.

\begin{figure}[hbt]
  \includegraphics[width=0.8\textwidth]{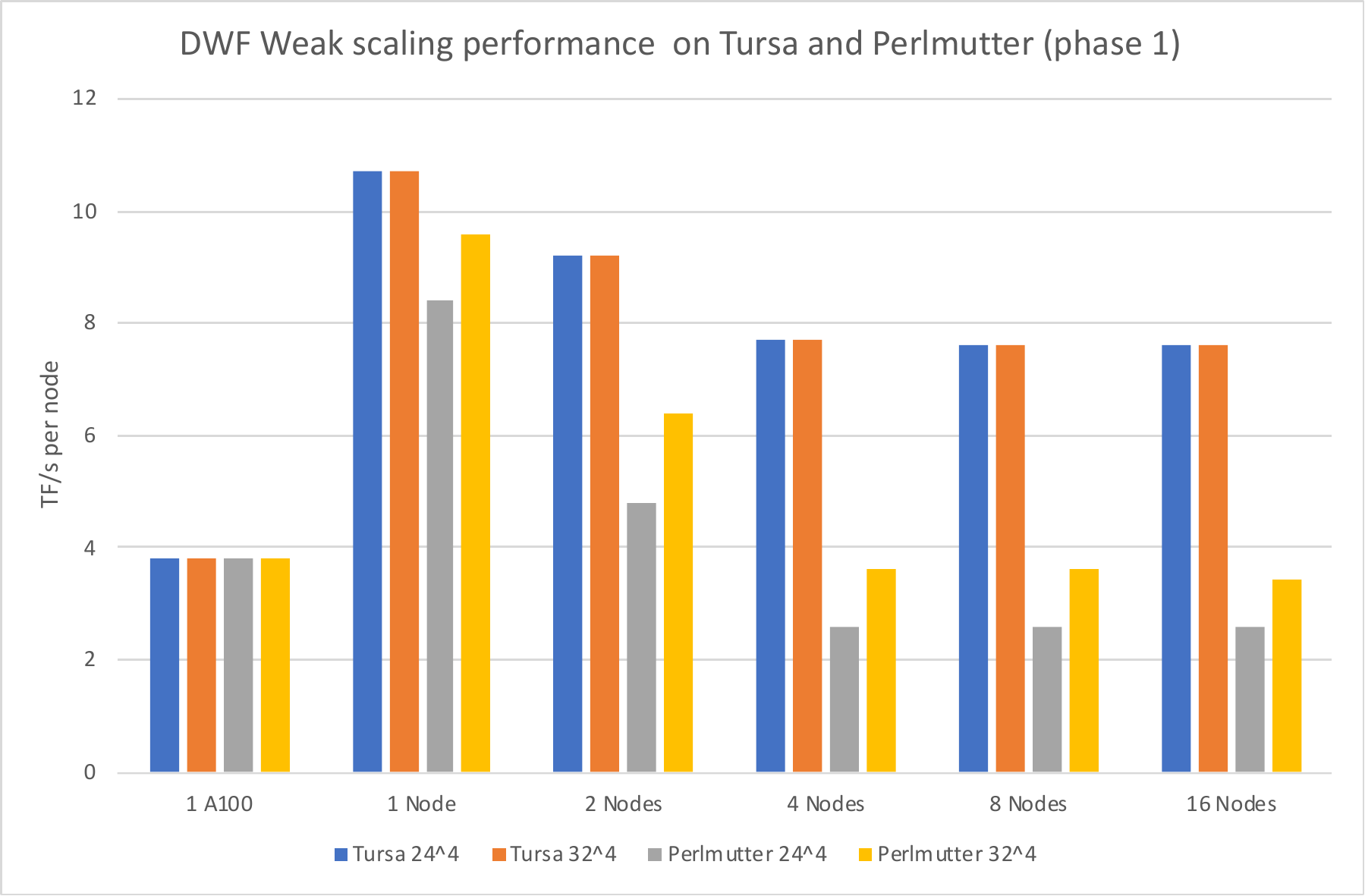}
  \caption{\label{perf}
    We show the performance per node on two recent systems comprising 4 x Nvidia A100 GPUs per node.
    The Atos Sequana ``Tursa'' system in Edinburgh (an identical technology to the Juelich Booster system),
    and the phase one Perlmutter system at NERSC, LBNL. The Perlmutter should be upgraded in phase 2 and
    is anticipated to give significantly upgraded performance. With current GPU's a ratio of 200Gbit/s interconnect per
    4TF/s seems a sweet spot.
  }
  \end{figure}

\begin{figure}[hbt]
  \includegraphics[width=0.8\textwidth]{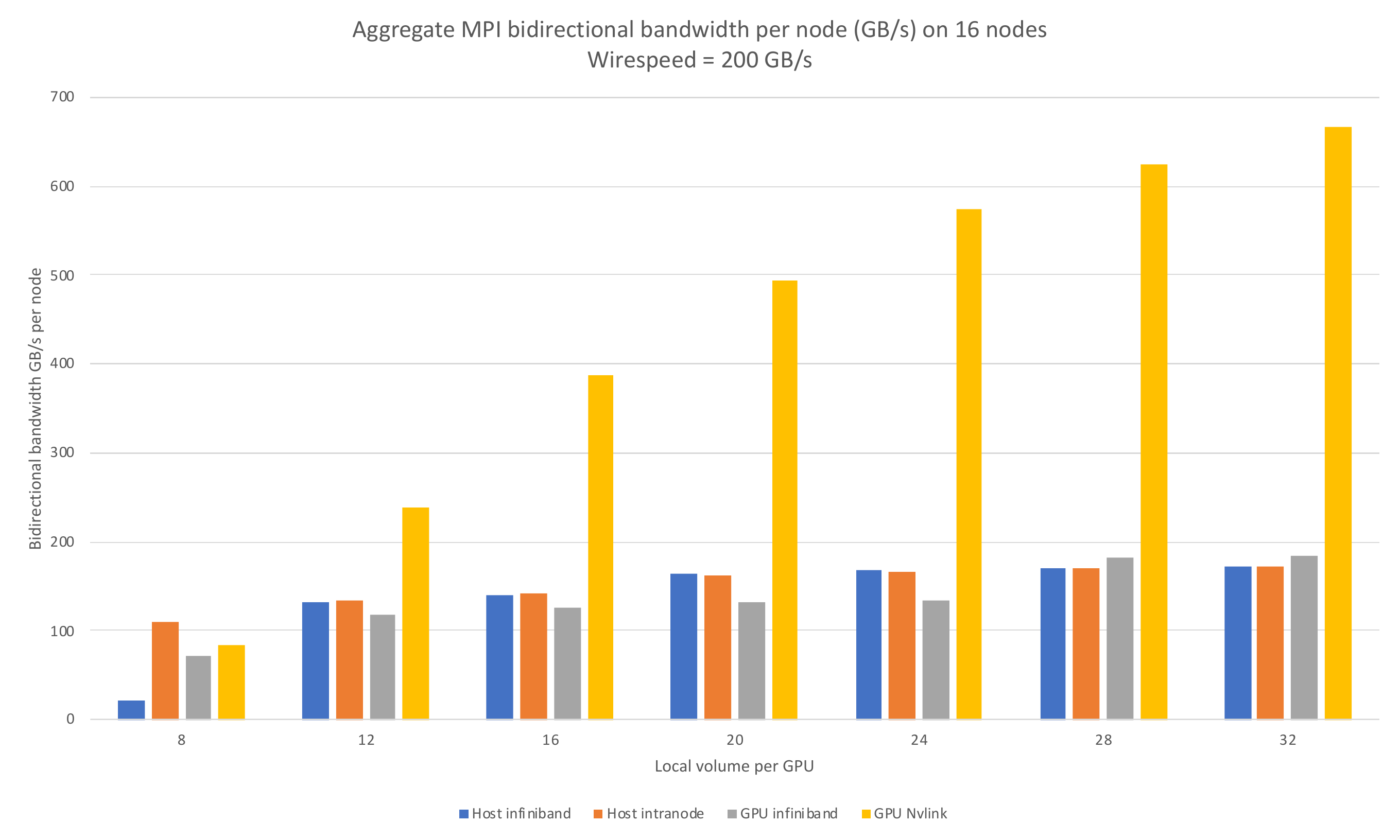}
  \caption{\label{network}}
  We show the GPU-GPU interconnect performance of four 200Gbit/s Mellanox infiniband cards on Tursa.
  The peak bidirectional bandwidth is 4 x 2 x 200 Gbit/s (and so 200 GB/s). Over 90\% of this speed
  is delivered using MPI to access GPU Direct RDMA bettween GPUs. NVlink performance interior to the node
  is excellent.
  \end{figure}

Figure~\ref{perf} displays the weak scaling of Grid on the Edinburgh ``Tursa'' and NERSC Perlmutter (phase 1) systems.
Phase 2 will shortly upgrade the network.
We see that Perlmutter is currently network limited but that Tursa has a balanced network provisioning that allows
good weak scaling at volumes per GPU of $24^4$ and above. The network performance of the ATOS systems is shown
in a detailed microbenchmark in figure~\ref{network}.

{\bf HIP}
We have run Grid but not yet fully optimised on the Spock system comprising 4 MI-100 AMD GPUs in ORNL.
We obtain 1.3TF/s per GPU and 4TF/s on one four-GPU node. We have been advised that performance patches
from AMD will increase this, perhaps to 1.8TF/s on MI-100. The Frontier system will install substantially faster MI-250
GPUs that the MI-100, and so we hope that a final configuration Frontier node will deliver a similar performance to a Tursa node.

{\bf SYCL}
We have run on Intel DGX and Arctic Sound GPUs obtaining expected performance consistent with the available
memory bandwidth. We used a mixture of the pure SYCL 2020 standard for Grid but dropped to the ``Level Zero'' vendor
specific API to access GPU-GPU copies within a node.

\begin{figure}[hbt]
  \includegraphics[width=0.75\textwidth]{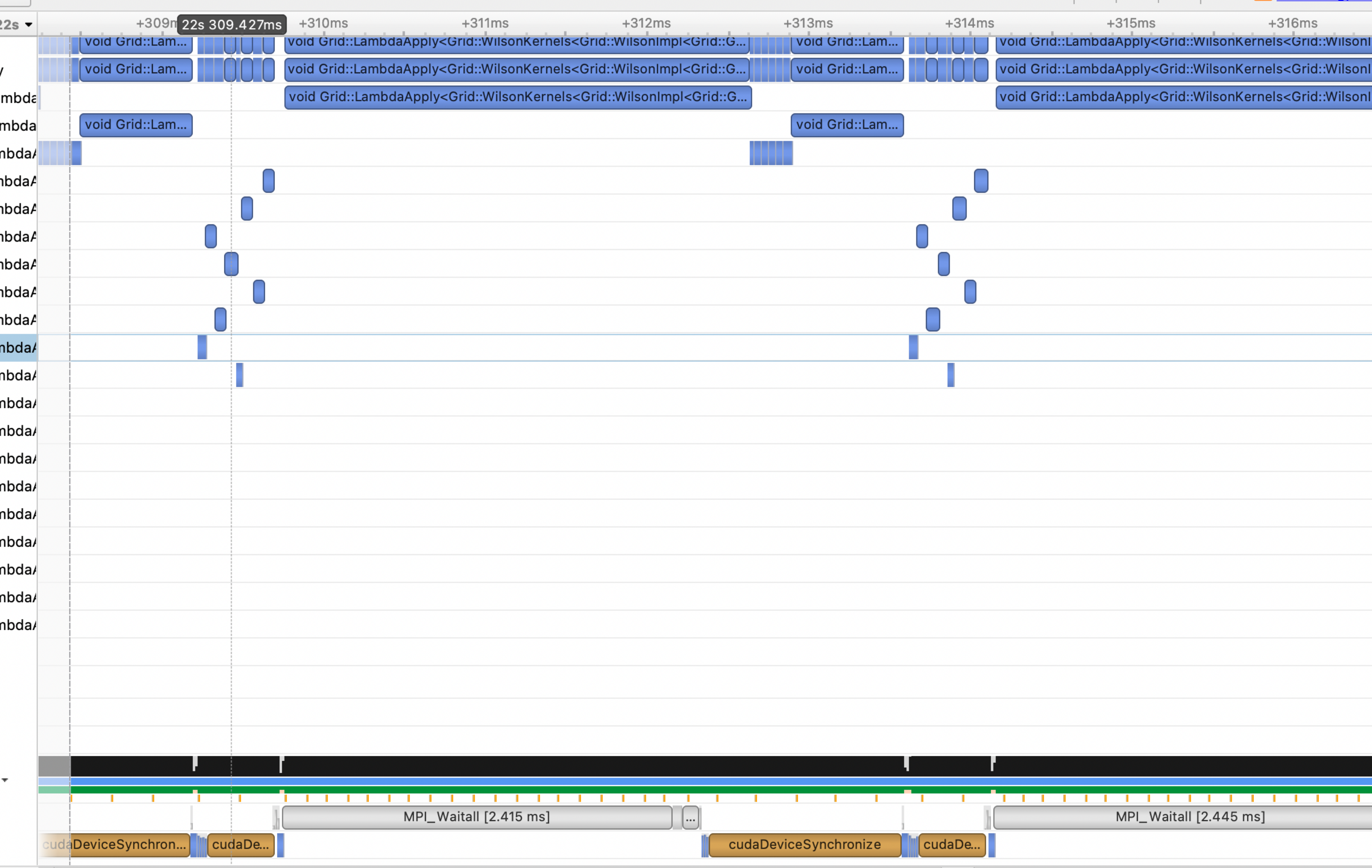}
  \caption{\label{nsys}
    We show the CUDA Nsight-sys profile of our code running on node zero of a 16 node job on Tursa.
    The communication and computation are perfectly overlapped and this system is well balanced for QCD.
    After continued optimisation of our code 16 multi-GPU nodes using 64 GPU's deliver that same performance
    as 1024 (substantially cheaper) nodes of the previous system it replaces.
    All kernels in the sequence (including face assembly) have been profiled and verified to obtain around 80\% of
    the peak memory bandwidth. 
  }
  \end{figure}

\begin{figure}[hbt]
  \includegraphics[width=0.8\textwidth]{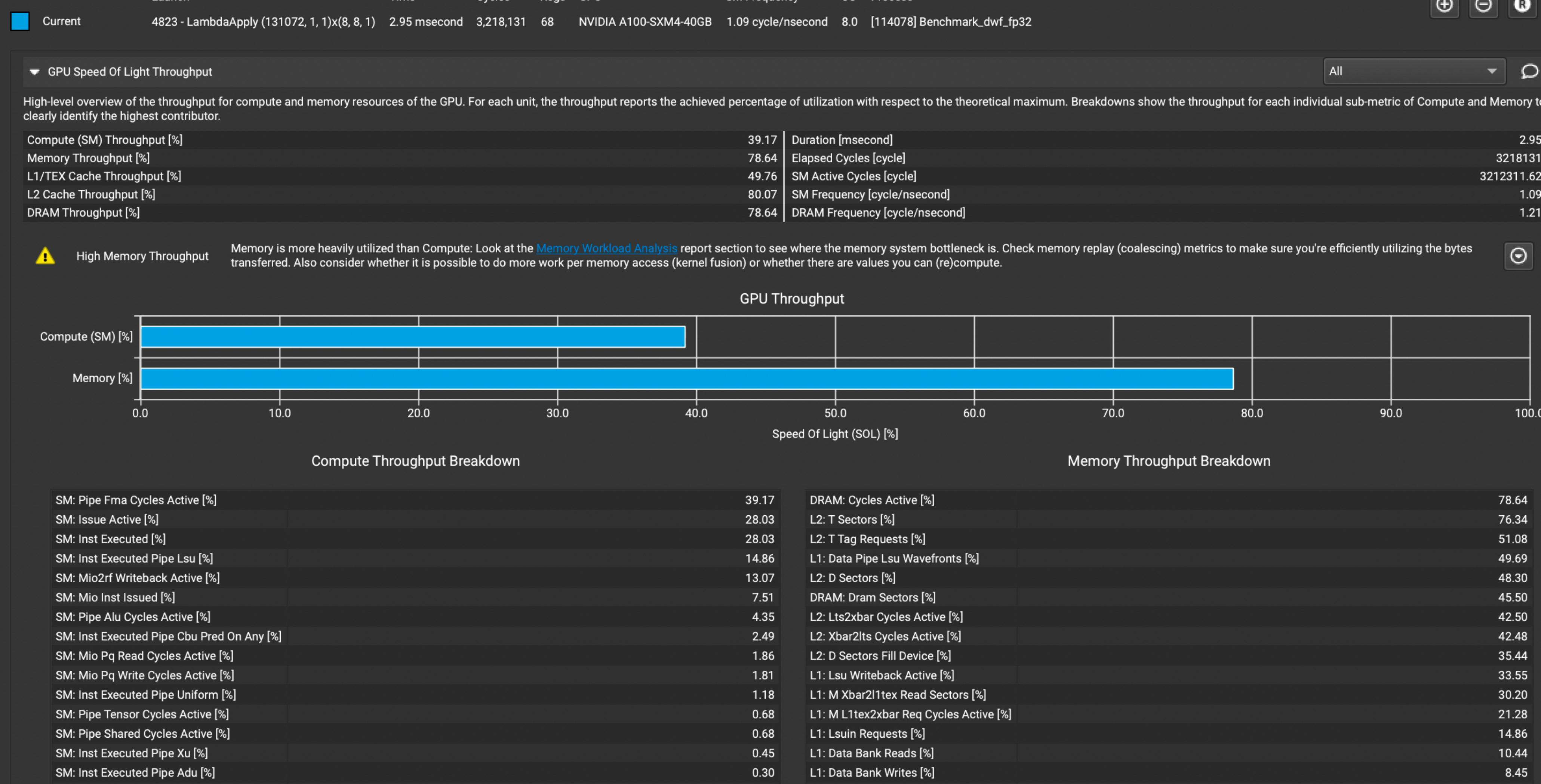}
  \caption{\label{ncu}
    We show the CUDA Nsight-compute profile of the main kernel DWF code.
    All kernels in the sequence (including face assembly) have been profiled and verified to obtain around 80\% of
    the peak memory bandwidth. This kernel is typical, but as it is floating point rich it is also seen to obtain 39\%
    utilisation of the floating point pipeline and a high fraction of the available cache bandwidth.
    Communication and computation are being efficiently overlapped while this kernel runs.
  }
  \end{figure}

\section{Conclusion and outlook}

Grid has been substantially reengineered to support both SIMD CPU and SIMT GPU execution models.
Of the planned Perlmutter, Frontier and Aurora systems in the US DOE open science roadmap, all of them
have distinct vendor native programming environments. Regardless Grid now supports all of these and is expected
to deliver good single GPU performance on each. Further, intranode communication is supported using direct
vendor provided GPU-GPU copy functionality and so is not dependent on an efficient MPI implementation.

On the Nvidia platforms the software has been profiled and demonstrated to saturate available memory bandwidth in all kernels
involved in the DWF Dirac operator and around 40\% of floating pipeline usage on the node local ``Wilson'' matrix.
The code is therefore provable optimal. Further excellent scaling is seen both within a node and on the Edinburgh and Juelich
systems across multiple nodes with near perfect overlap of communication and computation.

\section{Acknowledgements}

PB has been supported in part by the U.S. Department of Energy, Office of Science, Office of Nuclear Physics under the Contract No. DE-SC-0012704 (BNL). P.B. has also received support from the Royal Society Wolfson Research Merit award WM/60035. A.Y. has been supported by Intel.
A.P. is supported in part by UK STFC grant ST/P000630/1. A.P. also received funding from the European Research Council (ERC) under the European Union’s Horizon 2020 research and innovation programme under grant agreements No 757646 \& 813942.

\end{document}